# Density Scaling of the Dynamics of Vitrifying Liquids and its Relationship to the Dynamic Crossover

C. M. Roland[1]

*Naval Research Laboratory, Code 6120, Washington, DC 20375-5342*

and

R. Casalini[2]

*Naval Research Laboratory, Code 6120, Washington, DC 20375-5342*

*George Mason University, Fairfax, VA 22030*

*(December 21, 2004)*

[1] roland@nrl.navy.mil                    [2] casalini@ccf.nrl.navy.mil

A central question concerning glass-formation has been what governs the kinetic arrest of the quenched liquid – cooling reduces the thermal energy which molecules need to surmount local potential barriers, while the accompanying volume contraction promotes molecular crowding and congestion (and thus altering the potential). Recent experimental findings have shown that both thermal energy and density contribute significantly to the temperature-dependence of vitrifying liquids. Herein, we show that the scaling (superpositioning) of the relaxation times near the glassy state, by expressing them as a function of temperature and the specific volume, leads to a modification of the usual fragility curves, whereby differences in the extent of departure from Arrhenius behavior can be rationalized. More intriguingly, the characteristic changes in the relaxation properties (i.e., the "dynamic crossover"), occurring well above the liquid-to-glass transition, are shown to be related to the same function that superposes the relaxation data.

**1. Introduction**

Glass-formation is both a modern technology (e.g., many plastics, as well as metallic glasses) and an ancient craft (vitreous silicate artifacts date to the early Bronze age). Spectacular changes in physical properties accompany vitrification, yet no obvious changes occur on the molecular level. This makes the glass transition an intriguing phenomenon, which remains a major unsolved problem in the study of condensed matter. Glasses can be formed, *inter alia*, by cooling or by the application of large hydrostatic pressure. Since simple liquids tend to crystallize under such conditions, usually they must be quenched in order to attain the glassy state. This is not the case for most polymers, however, which resist crystallization because of chemical irregularities in their chain backbones. (In fact, one of the oldest earthly glasses is amber, formed by polymerization and concomitant vitrification of tree resin.) Because polymers readily form glasses, they were the focus of many early investigations of the glass transition. From these studies, the free volume theory of polymer dynamics was developed, which posits chain motion to be governed by the availability of empty space in the vicinity of the chain segments.[1,2] Free volume theory has been able to describe many of the dynamic properties of polymers. This, along with the intuitive appeal of the concept, led to its becoming the cornerstone of polymer viscoelasticity.[3]

In almost parallel fashion, albeit to a lesser degree, studies of the dynamics of molecular (non-polymeric) glass-formers were carried out, with quite different conclusions reached. Local motion is assumed to depend on a molecule's ability to surmount local energy barriers on the potential energy landscape [4,5,6,7]. There is no explicit role for free volume, although obviously the local density influences barrier heights. Buttressed by molecular dynamics simulations [8,9,10], this interpretation of the glass transition has become transcendent [11], with volume effects often relegated to temperatures well above $T_g$, the domain of Mode Coupling Theory [12]. Some recent efforts incorporate density fluctuations directly into an activated dynamics approach [13,14].

The contribution of density to the temperature dependence of relaxation times (or the viscosity, $\eta$, which is roughly proportional to $\tau$) can be quantified, provided that both the temperature- and density-dependences of the relaxation times are measured. We have been carrying out such experiments, using dielectric spectroscopy at varying hydrostatic pressure, on both molecular and polymeric glass-formers. The very general result is that volume effects cannot be neglected; indeed, in some cases their influence on the behavior can be more substantial than that of thermal energy.

## 2. Experimental

Our experimental approach is to determine the dielectric α-relaxation times as a function of both temperature and pressure. The complex permittivity is measured on samples contained between stainless-steel plates, with a separation ∼ 50 μm. The capacitor assembly is placed in a Manganin cell, where pressures as high as 1.4 GPa can be applied using a hydraulic pump, in combination with a mechanical intensifier (Harwood). The test sample and electrodes are isolated from the hydrocarbon pressurizing oil by Teflon. Dielectric spectra covering 10 decades of frequency are obtained using an Imass Time Domain analyzer ($10^{-4}$-$10^4$ Hz) and a Novocontrol Alpha Analyzer ($10^{-2}$-$10^6$ Hz). τ is obtained as the inverse of the frequency of the maximum in the dielectric loss associated with the α-process.

Pressure-volume-temperature measurements are carried out using a Gnomix instrument, based on the confining fluid technique, with the sample immersed in mercury. Pressures up to 200 MPa are applied at temperatures above the glass transition. The Tait equation of state is used for interpolation. In some case, data is presented herein which extends beyond the measured PVT range. While the Tait equation was used for extrapolation, we verified that an alternative equation state, the Padé equation[15], gives equivalent specific volumes. Note that all relaxation times presented herein are within the range measured dielectrically.

Most of the data presented herein came from our previous investigations using the techniques described above. For those, the original data and details of the experiments can be found in the published papers. The other data were measured using very similar methods.

## 3. Results and Discussion

To analyze the relaxation data in terms of the mutual interactions of the molecules is impossible without simplifying assumptions. Computer simulations of glass-formers[8,9,10] commonly rely on the Lennard-Jones 6-12 (LJ) potential, which has a repulsive term varying as $r^{-12}$, where $r$ is the separation between molecules. For dense liquids, a further simplification is to treat the weaker attractive forces as a spatially-uniform background (mean-field) potential, since long range forces on a molecule tend to average to zero. This emphasizes the dominant role of the short-range repulsive interactions for the local structure in non-associated liquids.[16,17,18] Of course, attractions affect the longer-ranged properties, such as the volume, and thus are necessary for an accurate equation of state; however, the arrangements and local dynamics are governed primarily by the repulsive interactions.[16,17]

Accordingly, recent simulation studies of the glass transition have employed an inverse power repulsive potential.[19,20,21] For the LJ fluid, this suggests a dependence of the relaxation times on the quantity $T^{-1}V^{-4}$ (where V is the specific volume $\propto r^3$), a scaling which turns out to be quite accurate for o-terphenyl (OTP).[22,23] The LJ potential is fairly "soft", and thus appropriate for a molecule such as OTP[10], comprised of planar phenyl rings. However, this $T^{-1}V^{-4}$ scaling fails to superimpose experimental results for most glass-forming liquids.

A generalized repulsive potential is the inverse power-law, $\varphi(r) \propto r^{-3\gamma}$, with the exponent $\gamma$ being material-specific.[24,25] Using this form, in principle all thermodynamic properties of the system, as well as local dynamic quantities such as structural relaxation, can be expressed in terms of the variable $T^{-1}V^{-\gamma}$. For the limiting case of hard spheres (for which volume dominates), $\gamma = \infty$, while for thermally activated dynamics, $\gamma = 0$. However, as noted above, a strictly repulsive intermolecular potential fails for properties such as the equation of state, due to the absence of an attractive term. It may also be inappropriate for associated liquids. However, our interest herein is restricted to the local dynamics. An inverse power-law potential (plus a mean field attractive term) was used by Shell et al.[19] in an energy landscape description of the glass transition. In this framework, they determined that the energies of the most and least stable inherent-structure configurations must scale with density as $V^{-\gamma}$.

In accord with such results, we have shown that the relaxation times for several glass-forming liquids and polymers, measured as isotherms at varying pressures and as isobars at varying temperatures, superpose when plotted versus $T^{-1}V^{-\gamma}$, where $\gamma$ is a material-specific constant.[26] We find $0.1 < \gamma < 9$ for glass-formers encompassing van der Waals molecules, associated liquids, and polymers.[26] The smaller values for $\gamma$ correspond to hydrogen-bonded liquids, in which volume effects are negligible. For small values of $\gamma$, the assumption that the repulsive interactions are short-range breaks down, and the superpositioning of $\tau$ is only phenomenological, albeit still useful.

An alternative linear scaling has been proposed, in which relaxation times are superposed by plotting versus $V^{-1} - V_0^{-1}$, in which $V_0$ is a constant. The quantity $TV^{\eta}$ can indeed be approximated by a linear function over a sufficiently narrow range of data, especially for small variations of the specific volume. However, over large ranges of $T$ and $V$, a linear approximation breaks down. This is demonstrated in Fig. 1 for propylene carbonate, using data from refs. 27 and 28. Similar failures of a linear function are seen for liquids having large values of $\gamma$ [29].

One implication of the scaling is that relaxation times can be depicted using an extended fragility plot, in which the abscissa $T^{-1}V^{-\gamma}$ is normalized by its value at $T_g$. [26] In Fig.2 we show such a modified fragility plot for 11 different liquids and polymers: 1,1'-di(4-methoxy-5-methylphenyl)cyclohexane (BMMPC) [28,30] 1,1'-bis(p-methoxyphenyl)cyclohexane (BMPC) [28,30,31]; cresolphthalein-dimethylether (KDE) [28,32]; $o$-terphenyl/$o$-phenylphenol (2/1) mixture (OTP/OPP) [33]; salol [34,35]; phenolphthalein-dimethylether (PDE) [36,37]; polymethyltolylsiloxane (PMTS) [38] ; propylene carbonate (PC) [27] ; poly(phenylglycidylether)-co-formaldehyde (PPGE) [39]; 1,2 polybutadiene (1,2-PB) [40]; sorbitol [41] . The slopes at $T_g$ correlate with the magnitude of $\gamma$, however $\gamma$ itself is not related to the isobaric fragility, as shown in ref.26. The problem is that, unlike $\gamma$, the isobaric fragility convolutes T and V effects. However, the slopes at $T_g$ do correlate with the isochoric fragility as discussed in ref. [42]. In Figure 3, we plot (in the inset) $\gamma$ versus the ratio of the isochoric and isobaric activation enthalpies, $E_V/E_P$, for 18 glass-formers including the 11 reported in Fig.2 together with: glycerol [43, 44], polypropylene glycol ($M_W$=4 kg/mol) (PPG4000) [45], polyvinylmethylether (PVME) [46],polyvinylacetate (PVAc) [47,48], diglycidylether of bisphenol A (DGEBA) [49], 1,4-polyisoprene (1,4-PI) [50], OTP [23], polymethylphenylsiloxane PMPS [51]. This activation enthalpy ratio varies between zero (volume-dominated dynamics) to unity (thermally-activated dynamics).[52] As expected, the two quantities inversely correlate, since larger $\gamma$ implies a stronger effect of $V$ on the intermolecular barriers to local rearrangements.

If the relaxation time is a function of the quantity $TV^{\gamma}$, it is easy to show that

$$\left.\frac{E_V}{E_P}\right|_{T_g} = \frac{1}{1+\gamma T_g \alpha_P(T_g)} \qquad (1)$$

where $\alpha_P$ is the isobaric thermal expansion coefficient. In Fig. 3, the value of $\gamma$ calculated from eq. 1 is plotted versus the $\gamma$ which gives superpositioning of the $\tau(T,V)$ data. The agreement between the calculated and directly determined $\gamma$ is quite good, especially in light of the fact that the data come from different sources, employing different techniques, and of course this agreement corroborates the scaling. Note that since $E_V / E_P = 1 / \left( 1 - \alpha_P / \alpha_\tau \right)$ [46], where $\alpha_\tau$ is the thermal expansion coefficient at constant $\tau$, eq. 1 yields the relation, $\gamma = -1 / \left( T_g \alpha_P (T_g) \right)$.[53] The implied constancy of the product of the isobaric thermal expansion coefficient and the glass transition temperature is known empirically as the Boyer-Spencer rule.[54]

While the results in Fig. 3 give a quantitative accounting of the effects of specific volume and temperature in the vicinity of $T_g$, upon cooling through a higher temperature, $T_c$ ($\sim$ 1.2 to 1.4 $\times$ $T_g$), characteristic changes in the dynamics of a liquid are observed: (i) Departures in the proportionality of the viscosity to the translational diffusion constant (the Stokes-Einstein relation) and to the orientational relaxation time (Debye-Stokes-Einstein relation)[55,56,57,58]. (ii) The relaxation function broadens markedly [59,60], while the dependence of the relaxation time on the configurational entropy departs from the form of the classical Adam-Gibbs theory [61,62]. (iii) The glass transition relaxation bifurcates, with the emergence of a Johari-Goldstein secondary relaxation process [63,64]. (iv) The derivative of $\tau$ with respect to T changes slope, indicating a change in the temperature-dependence [37]. (vii) A change in the temperature dependence of the dielectric strength [65]. The cause of these characteristic changes in properties, at a temperature well above the glass transition, is unclear. According to MCT [12], $T_c$ is associated with a change from liquid-like to solid-like dynamics (or to a divergence). Similarly, a free volume model [2] predicts a characteristic temperature at which continuity of liquid-like cells is attained, and experimentally, this temperature corresponds to $T_c$ [66]. Thus, the crossover appears to denote the onset of "caged dynamics". Since the dynamic behavior of the liquid is changing at this temperature, the $T_c$ phenomenon is referred to as the dynamic crossover. An intriguing aspect of experimental work is that while $T_c$ is pressure-dependent, the value of $\tau$ (or $\eta$) at $T_c(P)$, is constant for a given liquid [67,68]. Since the value of $\tau_c$ ($= \tau(T_c)$) does not vary greatly among most liquids, it has even been referred to as a "magic relaxation time" [69].

Understanding the mechanisms underlying the dynamic crossover of liquids, which occurs so far above their glass transition temperature, is an important aspect of solving the glass transition problem. If $\tau_c$ is indeed a material constant, we predict that the product $T_c^{-1} V_c^{-\gamma}$ should also be constant. This leads to an equation for the characteristic temperature

$$T_c = \frac{1}{\alpha_P(T_c)\gamma} \left( \left. \frac{E_P}{E_V} \right|_{T_c} - 1 \right) \qquad (2)$$

Eq. 2 reveals that the temperature of the crossover in glass-forming liquids and polymers is directly related to the relative effect that temperature and volume have on the dynamics (as reflected in the magnitude of $E_V/E_P$). In Table 1, we compare for several glass-formers the characteristic temperature calculated from eq. 2 (with no adjustable parameters) to the experimentally determined $T_c$. The agreement is quite satisfactory, as evident also from Figure 4 where $T_c$ calculated using eq.(2) are plotted versus the average value of $T_c$ from various experimental determinations. These results (Table 1 and Fig. 4) also corroborate the assumption underlying eq. 2 that $\tau_c$ is a material constant, independent of $T$ and $P$. This invariance of $\tau_c$ has previously been observed directly for several materials, from measurements at elevated pressure [42,67,68].

**4. Conclusions**

In summary, both temperature and volume exert a significant influence on the dynamics of supercooled liquids, and their relative effects near $T_g$ are experimentally quantifiable. The superpositioning of the relaxation times enables relaxation properties for diverse glass-formers to be expressed as a single function of $T$ and the specific volume $V$. This provides a more transparent comparison of departures from Arrhenius behavior (i.e., fragility) for different materials. Whereas the conventional fragility plot gives no explicit consideration to any dependence of the activation energy on $V$, the present modification (Fig. 1) offers a clear separation between the effects of T and V on τ. For example, this approach distinguishes between materials having strong interactions (hydrogen-bonded) and those more weakly bonded (van der Waals), although these may have comparable fragilities.

We also show that from the condition of constancy of the characteristic relaxation time τ$_c$, the scaling exponent $\gamma$ is directly related to the temperature, $T_c > T_g$, at which characteristic changes occur in the dynamic properties of the liquid. This means that molecules are responding to the conditions responsible for glass formation well before vitrification *per se* commences. Thus, further progress in understanding the glass transition can be made by focusing on the precursor events transpiring well above $T_g$. Furthermore, the scaling, in combination with the constancy of $\tau_c$, leads to the conclusion that the ratio of the mean squared displacement to the nearest-neighbor spacing is a constant at the crossover for any $T$ and V (*viz.* the Lindemann law).[24] Similar ideas have been expressed previously regarding the crossover [69], and are in agreement with the interpretation of the crossover as a signature upon cooling of the beginning of caged dynamics (onset of strong intermolecular cooperativity).

Finally, the experimental verification of eq. 1, together with the apparent linear behavior in Fig.1 on approaching $T_g$ ($\tau \sim 10^2$ s) suggests that the marked slowing down of molecular motions on approaching the glass transition can be described as a thermally activated process with a progressively increasing activation energy ($E_a \sim V^{-\gamma}$). Consequently, if this interpretation is correct, there would be no need to invoke a divergence at some thermodynamic transition below $T_g$.

**Acknowledgement.**

This work was supported by the Office of Naval Research.

**Figure captions**

Figure 1. Dielectric α-relaxation times for PC, as a function of (a) $T^{-1}V^{-3.8}$ and (b) the linear approximation $T^{-1}(V^{-1} - V_0^{-1})$. For the latter, neither superposing the data at low temperatures ($V_0 = 0.966$) or high temperatures ($V_0 = 1.053$) yields satisfactory scaling over the entire range of the measurements. The experimental data are from refs. 27 and 28.

Figure 2. Modified fragility plot, $\log(\tau)$ as a function of $T^{-1}V^{-\gamma}$ normalized by its value at the glass transition ($\tau = 10^2$ s) $T_g^{-1}V_g^{-\gamma}$. The parameter γ was obtained from the superpositioning of the relaxation time data. For clarity, only data at atmospheric pressure are shown here, since those at high pressure fall onto the same curve. The data are reported for eleven representative glass-formers: BMMPC γ=8.5 (▲); BMPC γ=7 (▽); KDE γ=4.5 (◁); OTP/OPP γ=6.2 (▶); salol γ = 5.2 (○); PDE γ=4.5 (✕); PMTS γ=5 (△); PC γ=3.8 (□); PPGE γ=3.5 (+); 1,2-PB γ=1.9 (☆); sorbitol γ=0.13 (▼).

Figure 3. The $\gamma_{calc}$ parameter calculated using equation $\gamma^{-1} = \alpha_\tau T_g$, versus the parameter γ obtained from the scaling plot for different glass formers. The dotted line is the best fit straight line to the data, while solid line represents $\gamma_{calc}$ = γ. The inset shows the ratio of the isochoric and isobaric activation enthalpies as a function of γ. 1 - ideal T-activated; 2 - sorbitol; 3 - glycerol; 4 - 1,2-PB; 5 – PPG4000; 6 - PVME; 7 – PVAc; 8 - DGEBA; 9 – 1,4-PI; 10 - PPGE; 11 - PC; 12 - OTP; 13 – PDE; 14 – KDE; 15 – PMTS; 16 – salol; 17 - PMPS; 18 - BMMPC; 19 - BMPC.

Figure 4. Dynamic crossover temperature calculated from γ (eq.(2)) for the change in dynamics occurring above $T_g$, plotted versus the average of the experimental values of $T_c$ reported in Table 1.

**Table captions**

Table 1. Values of $T_c$ as estimated from experiments compared with the values of $T_c$ calculated using eq.(2). The bold values came from dielectric relaxation data.

| | $T_c[K]$ experimental | $T_c [K]$ calculated |
|---|---|---|
| *PC* | 176 [70], 187 [71], **187** [72] 196 [73], **200** [3737] **189** [42] | 191 |
| *OTP* | 285 [74], 290 [75], 293 [76], **290** [31] | 303 |
| *Salol* | 256 [77], 263 [78], 266 [79], 275 [80], **265** [37], **253** [42] | 264 |
| *PDE* | **325** [37], **319** [81] **322** [42] | 300 |
| *BMPC* | **270** [31] | 290 |
| *BMMPC* | **320** [28], **315** [42] | 299 |
| *PVAc* | **383** [82] | 381 |

**Table 1**

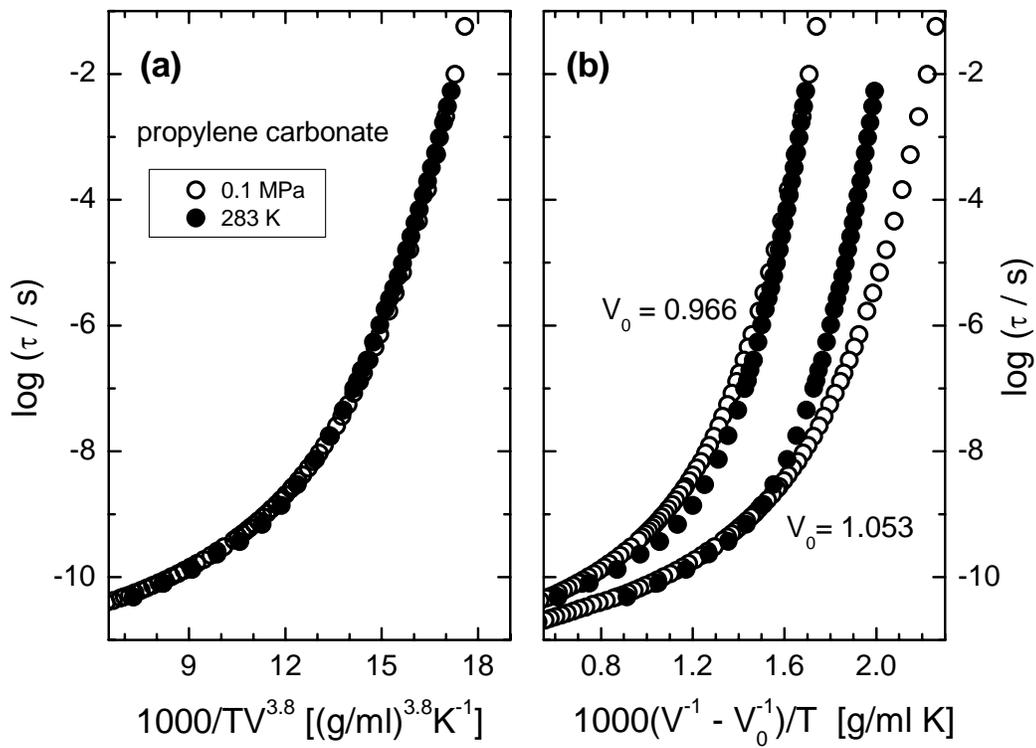

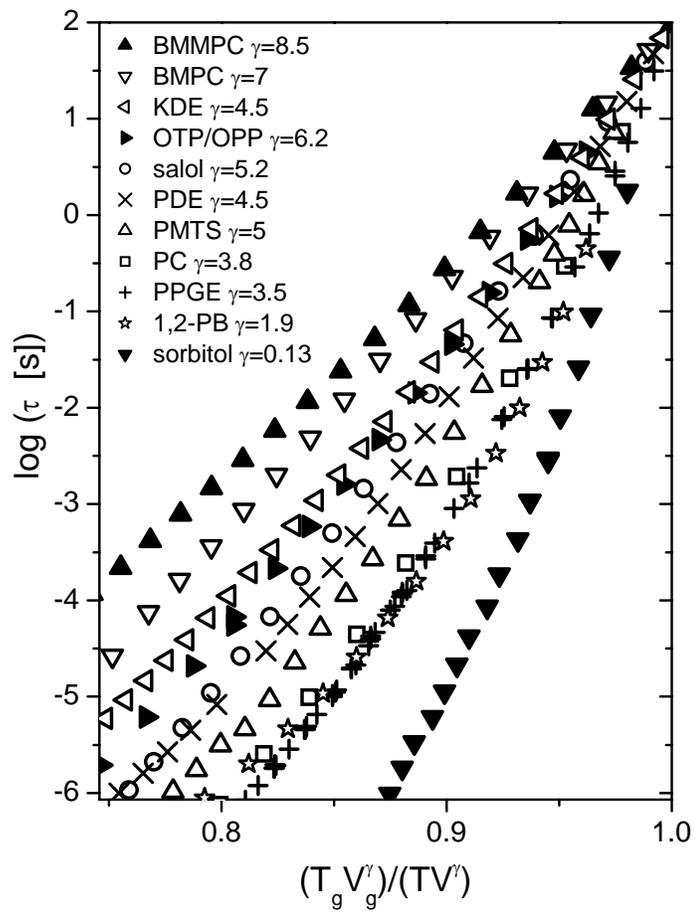

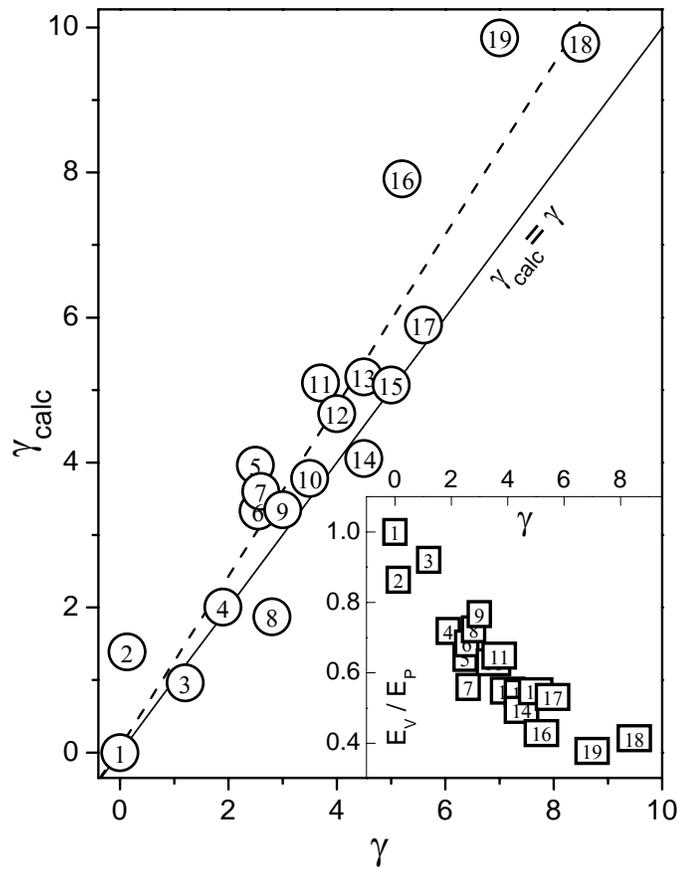

figure 3

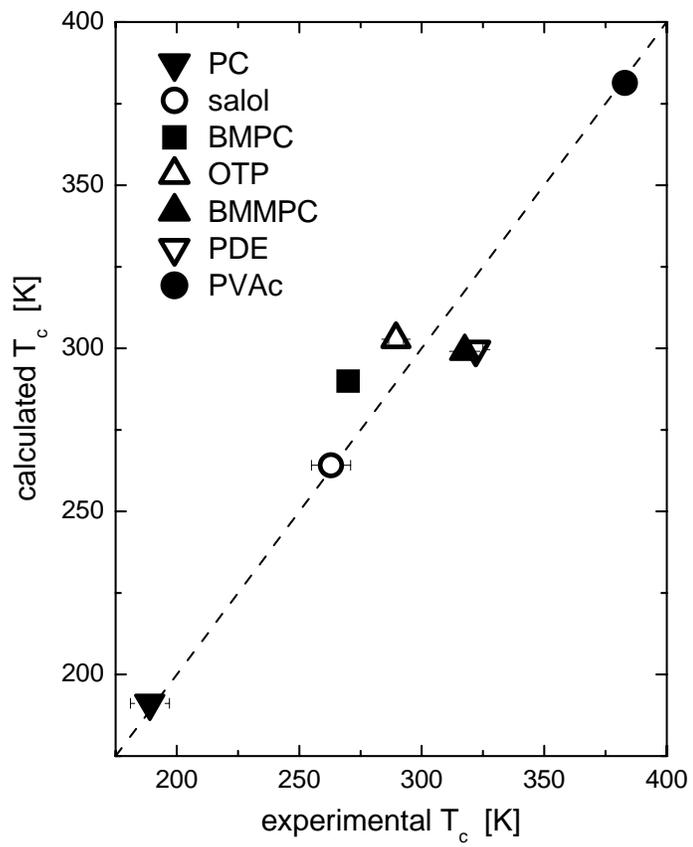